**Development of an expected possession value model to analyse team attacking performances in rugby league**

Running title: Expected possession value in rugby league

Thomas Sawczuk[1,2,3], Anna Palczewska[1], Ben Jones[2,3,4,5]

[1] School of Built Environment, Engineering and Computing, Leeds Beckett University, Leeds, United Kingdom

[2] Carnegie Applied Rugby Research (CARR) Centre, Carnegie School of Sport, Leeds Beckett University, Leeds, United Kingdom

[3] England Performance Unit, The Rugby Football League, Red Hall, Leeds, United Kingdom

[4] Leeds Rhinos Rugby Club, Headingley Carnegie Stadium, Leeds, United Kingdom

[5] School of Science and Technology, University of New England, Armidale, New South Wales, Australia

Abstract words: 198

Manuscript words: 3,959

Figures: 4

Tables: 0

**Abstract**

This study aimed to provide a framework to evaluate team attacking performances in rugby league using 59,233 plays from 180 Super League matches via expected possession value (EPV) models. The EPV-308 split the pitch into 308 ~5m x 5m zones, the EPV-77 split the pitch into 77 ~10m x 10m zones and the EPV-19 split the pitch in 19 zones of variable size dependent on the total zone value generated during a match. Attacking possessions were considered as Markov Chains, allowing the value of each zone visited to be estimated based on the outcome of the possession. The Kullback-Leibler Divergence was used to evaluate the reproducibility of the value generated from each zone (the reward distribution) by teams between matches. The EPV-308 had the greatest variability and lowest reproducibility, compared to EPV-77 and EPV-19. When six previous matches were considered, the team's subsequent match attacking performances had a similar reward distribution for EPV-19, EPV-77 and EPV-308 on 95 ± 4%, 51 ± 12% and 0 ± 0% of occasions. This study supports the use of EPV-19 to evaluate team attacking performance in rugby league and provides a simple framework through which attacking performances can be compared between teams.

## Introduction

In recent years, the growing availability of event level data in rugby league has led to an increase in research surrounding the characteristics of match winning rugby league performances (Kempton, Kennedy and Coutts, 2016; Woods, Sinclair and Robertson, 2017; Parmar *et al.*, 2018; Wedding *et al.*, 2020). These studies can broadly be split into two categories based on the inclusion of spatial data within their analyses (Kempton, Kennedy and Coutts, 2016), or not (Woods, Sinclair and Robertson, 2017; Parmar *et al.*, 2018; Wedding *et al.*, 2020). Spatial data provides valuable contextual information within the analysis of event level data by identifying the location of each event. Whilst studies not including spatial data provide valuable insights into potential match winning actions (Parmar *et al.*, 2018) or the classification of player positions (Wedding *et al.*, 2020), they do not account for some of the most valuable contextual information surrounding the location of the events analysed. Incorporating this spatial context into the analysis of a team's attacking performances could have a significant impact on tactical preparations for future matches and therefore provides a valuable avenue of research within rugby league.

Typically, spatial data has been included in analyses of team and player performance via expected possession value (EPV) models (Cervone *et al.*, 2016; Kempton, Kennedy and Coutts, 2016; Fernández, Bornn and Cervone, 2019). These models assign a value to every event or pitch location visited during a match based on the probability of scoring a goal, basket or try from that action/location within a given amount of time. When incorporating spatial data within EPV models, it is common to split the pitch into different zones, which pool data together. Discretising the pitch into these zone systems is computationally efficient and allows for improved generalisability to other samples. Nevertheless, selecting the correct zone system is fraught with difficulty: if the zones are too large, valuable data will be lost; if they are too small, the results of the analysis will not be generalisable (Bivand, Pebesma and Gomez-Rubio, 2013).

Within the EPV literature, two key methods have been used to select zone systems: fixed zone sizes (Kempton, Kennedy and Coutts, 2016); and the selection of zones based on likely shooting locations (Cervone *et al.*, 2016). Although it isn't possible to identify shooting locations within rugby league given a try can be scored across the width of the pitch, it may be possible to obtain a rugby league specific zone system by aggregating zones together based on the total value they generate during a match (i.e. their match return). This could allow zones which generate a similar match return to be aggregated, providing a more sport specific EPV zone system than the fixed size grid method previously considered (Kempton, Kennedy and Coutts, 2016). However, no study has yet compared the two methods of selecting zone sizes.

When comparing different zone systems, it is important that the criterion measure used is relevant to the intended purpose of the model. Within a model evaluating rugby league attacking performances, one of the most important elements is the reproducibility of performances between fixtures (Shafizadeh, Taylor and Peñas, 2013). EPV models quantify attacking performances through the individual zones' match returns. Therefore, to understand the reproducibility of attacking performances from the perspective of an EPV model, it is necessary to evaluate how similar the match returns obtained by each zone are between fixtures. Completing such an analysis within EPV models using the previously published fixed zone size (Kempton, Kennedy and Coutts, 2016), a smaller fixed zone size and an aggregated zone system would therefore help to identify the most suitable zone system for use within rugby league. Furthermore, it would be useful to provide a methodology through which the EPV model could be used to identify the zones a team generates value from within their attacking performances. Such a framework was not provided by Kempton et al. (2016) in their previous study, limiting the use of the EPV model in practice.

As a result of the limitations surrounding EPV models within the sports science domain, the aim of this study was to produce a framework through which an EPV model could be used to analyse a team's attacking performances in rugby league. To achieve this aim, three key objectives were identified: a) generate three EPV models, two with fixed zone sizes of ~5m x 5m and ~10m x 10m (Kempton, Kennedy and Coutts, 2016), and one with aggregated zones based on the individual zones' match returns; b) identify the zone size which provided the greatest reproducibility of attacking performances between fixtures; and c) propose a framework through which an EPV model could be used in practice to evaluate a team's attacking performances.

## Methods

### *Sample*

Event level data were obtained from Opta (Stats Perform, London, UK) for all 180 matches of the 2019 Super League season. In total, 59,233 plays were analysed. Within this sample, 1,369 tries were scored (1,013 successful conversions, 356 unsuccessful conversions), 271 penalty goals were attempted (239 successful, 32 unsuccessful) and 89 drop goals were attempted (42 successful, 47 unsuccessful). Prior to analysis, ethics approval was provided by a University sub-ethics committee.

*Data pre-processing*

Event level data typically includes data surrounding the action completed (e.g. pass, tackle, try scored), the players involved, the time, and the set and play count. However, Opta only includes location data for the first action of each play, so only this information, provided as x and y co-ordinates, was used to provide spatial context within this study. Despite some variation being present in pitch sizes across the Super League, Opta standardises pitch dimensions to 68m x 120m through its coding software so these dimensions were used for this study.

For the two fixed zone size models, the 68m x 120m pitch was separately split into 336 ~5m x ~5m zones and 84 ~10m x ~10m zones. The zones inside the opposition team's try area were removed as obtaining the ball in these positions is implicitly of high value and would likely result in an attempt to ground the ball before the next play begins. This left 77 zones of ~10m x ~10m size for the previously published fixed zone size (EPV-77; (Kempton, Kennedy and Coutts, 2016)) and 308 zones of ~5m x ~5m size for the smaller fixed zone size (EPV-308). In both models, the widest zones were 1m narrower than any central zones. The use of these narrower zones was necessary as Opta only provides location data to the nearest metre. Consequently, splitting the 68m pitch width into equal 4.85m or 9.7m zones would provide no additional detail and would be much more difficult to understand in practice.

In preparation for the main analyses, attacking performances were split into attacking possessions. An attacking possession was determined to begin when a team obtained possession of the ball and ended when the team lost control of the ball, either due to an error, handover, field kick, penalty or drop goal attempt, or by scoring a try. Using this definition of an attacking possession, the 59,233 plays were split into 10,156 attacking possessions with a median length of 4 plays per attacking possession (range 1-26 plays). For every attacking possession, the zone providing the location at the beginning of each play was recorded so an attacking possession consisted of a sequence of movements between zones. At the end of each attacking possession, a reward was provided dependent on the outcome of the possession: converted try scored (+6); unconverted try scored (+4); penalty goal scored (+2); drop goal scored (+1); loss of possession or missed goal attempt (0). No intermediate rewards were provided.

*Calculation of fixed zone size EPV values*

To evaluate the EPV for each zone on the pitch within the fixed zone size systems, attacking possessions were considered as Markov Chains whereby the location of the ball on the pitch (i.e. its zone) at a given time within the possession was represented as an event. The value for each zone *s*, at a given time within the possession *t* was defined as:

$$V(s,t) = E[\sum_{k=0}^{\infty} \gamma^k R_{t+k+1} | S_t = s] \quad (1)$$

where V(s, t) is the estimated reward obtained from the zone *s*, at the time *t*, $R_u = R_{t+k+1}$ is the reward obtained at the time *u*, which is determined by the end of play *u-1* (e.g. if a converted try is scored at play u-1, $R_u = 6$), γ is the discount factor and *k* is a play within the attacking possession.

Subsequently, the overall return of any zone *s* after play *t* across the sample of attacking possessions was calculated as:

$$G(s,t) = \sum_{j \in A_{s,t}} \gamma^{\tau_j - 1 - t} R^j \quad (2)$$

where G(s, t) refers to the overall return for zone *s* after play *t* across the sample of attacking possessions; $A_{s,t}$ is a set of attacking possessions where the ball is at location *s*, in play *t*; $\tau_j$ is the play number within an attacking possession *j*, $R^j$ is the reward for the attacking possession *j*.

Finally, the expected value (EPV) of each zone *s* after time *t* was simulated using the Monte Carlo every visit algorithm as follows:

$$EPV(s,t) \approx \frac{1}{|A_{s,t}|} \mathrm{G(s,t)} \quad (3)$$

*Calculation of aggregated zone system*

To calculate the aggregated zone system (EPV-19), EPV-308 zones were grouped together based on their match return. The match return ($G_m$(s,t)) for zone *s* in match *m* was calculated using Equation (2). Each zone's match return was summed at a column level to provide the column match return, and at a row level to provide the row match return. This resulted in 14 columns of ~5m width and 22 rows of 5m height. Visual inspection of the initial columns showed that there were no differences in the match return at the 5m level symmetrically (i.e. the two widest columns had very similar values, as did the two most central and all others in between), so these values were aggregated to form seven ~10m columns. Similarly, visual inspection of the data showed the rows could be aggregated at a 10m level, similar to Kempton et al. (2016), resulting in eleven 10m rows.

Following this initial aggregation, linear mixed models were used to evaluate whether the columns or rows could be further combined. In separate models, the column match return and row match return were added as dependent variables, with team and fixture ID added as random effects. To evaluate for differences in their respective match returns, column and row indexes were added to their respective models as categorical fixed effects. Minimal effects testing (Murphy and Myors, 1999) was used to determine whether two columns or rows could be combined against a smallest effect size of interest of 1 unit of match return. Statistical significance for the difference was recognised when $P < 0.05$. If the difference between two columns or rows was statistically significant, they remained separate; otherwise, their match returns were averaged and compared to the next column or row. This iterative process was conducted independently for the columns and rows, resulting in six columns and four rows. Finally, the columns and rows were combined to create the EPV-19 zone system. The first row (-10m to 10m) was not split into six columns due to both the high variability of the zone values within the row, and the infrequency with which they were visited relative to the other rows. The values of the EPV-19 zones were calculated as a weighted average of the values of the EPV-308 zones they were composed of. Figure 1 depicts the similarities and differences between the EPV-308, EPV-77 and EPV-19 in the 30m closest to the opposition try line.

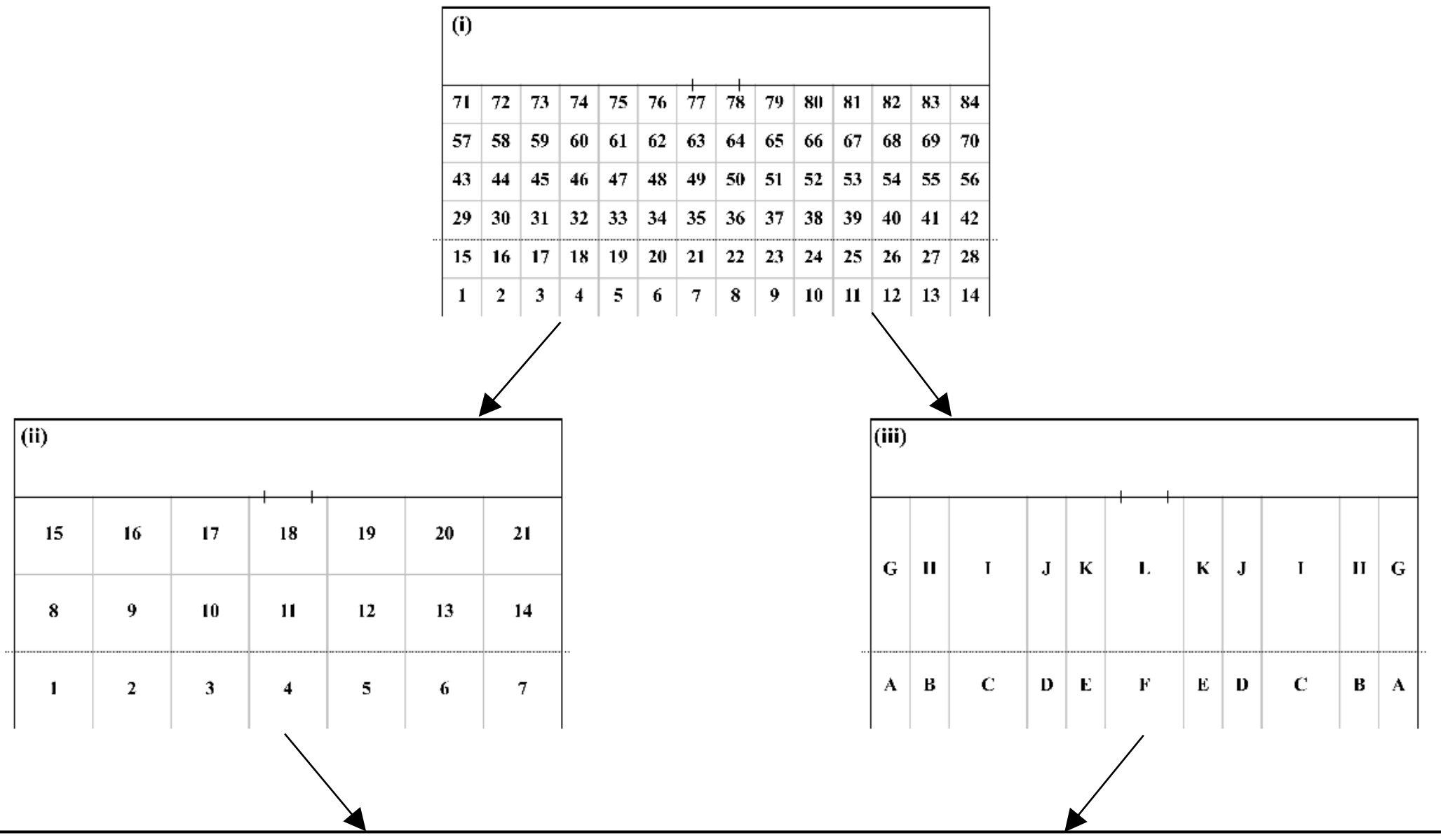

EPV-77 zone 4 and EPV-19 zone F are both weighted averages of EPV-308 zones 7, 8, 21 and 22, so they have equivalent EPV values.

EPV-19 zone I is the weighted average of zones 35, 36, 49, 50, 63, 64, 77 and 78. EPV-77 zone 11 (weighted average of zones 35, 36, 49 and 50) or EPV-77 zone 18 (weighted average of zones 63, 64, 77 and 78). Consequently, these three zones have different EPV values.

**Figure 1: Depiction of similarities and differences between EPV-308 (i), EPV-77 (ii) and EPV-19 (iii) in the 30m closest to the opposition try line. Each zone from the EPV-77 and EPV-19 is a weighted average of the EPV-308 zones they are composed of. Dotted line represents the opposition 20m line.**

*Evaluating the reproducibility of attacking performances*

To evaluate the reproducibility of attacking performances between fixtures, the match returns from a specified number of previous fixtures were compared to the subsequent fixture. The previous 1-10 fixtures were considered with every possible combination of fixtures evaluated for each team within the Super League. Consequently, 28 comparisons were possible for each team when 1 previous fixture was used (e.g. match 1 vs 2, match 2 vs 3, match 3 vs 4 etc), 27 comparisons were possible when 2 previous fixtures were used (e.g. match 1 and 2 vs 3, match 2 and 3 vs 4 etc), through to 19 comparisons when 10 previous fixtures were utilised.

The Kullback-Leibler (KL) Divergence (Kullback and Leibler, 1951) was used to calculate the similarity between the reward distribution in the subsequent match *i* and the reward distribution in the *k* previous match(es). The reward distribution for zone *s* in match *mi* ($PG_{mi}(s)$) was calculated via the equation:

$$PG_{mi}(s) = \frac{G_{mi}(s)}{\sum_{s=1}^{S} G_{mi}(s)} \qquad (4)$$

The reward distribution for zone *s* in *k* matches prior to match *i* was calculated as:

$$PG_{M}(s) = \frac{\sum_{k=1}^{k=i-1} G_{mk}(s)}{\sum_{k=1}^{k=i-1} \sum_{s=1}^{S} G_{mk}(s)} \qquad (5)$$

where $PG_M(s)$ refers to the reward distribution obtained by zone *s* across *M* fixtures, $G_m(s)$ refers to the match return obtained by zone *s* in match *m* and *S* refers to the set of all zones within the EPV model.

The KL Divergence is a measure used in information theory and provides an understanding of the similarity between two distributions of values. It is an unbounded measure, where a value of 0 indicates two distributions are perfectly matched, but a value of infinity indicates that there is no relationship between the two distributions. A value of infinity typically occurs when a zone within the true distribution has a high value, but the corresponding zone in the approximating distribution has a low value. In this study, the subsequent fixture's reward distribution was used as the true distribution and the previous fixture(s)'s reward distribution was used as the approximating distribution. The percentage of non-infinity values was used to provide an understanding of how many subsequent fixtures' attacking performances were reproducible given the reward distributions of the previous fixtures. These values are provided as a mean and standard deviation across the twelve Super League clubs.

*Identifying zones important to attacking performances*

A simple framework through which the EPV models could be used in practice to evaluate a team's attacking performances was provided using z-score analysis. The team's reward distribution across the whole 2019 Super League season was calculated for the EPV-19 via Equation 5. Z-score analysis of the reward distributions was used to calculate a standardised value evaluating how much or little a team relied on a specific zone when attacking compared to the average dependence across all teams in the Super League. Values of +1 and +2 z-scores were chosen to represent high and very high dependence on zones, values of -1 and -2 were used to represent low and very low dependence.

All analyses were conducted using bespoke Python scripts (Python 3.7, Python Software Foundation, Delawere, USA) or via Proc Mixed (SAS University Edition, SAS Institute, Cary, NC)

## Results

### *Generation of EPV models*

Figure 2 illustrates the three EPV models (EPV-308, EPV-77 and EPV-19). There is a general trend that the closer the zone is to the opposition try line, the more valuable it is. Similarly, central zones are more valuable than wider zones as indicated by the darker colours in these areas. Furthermore, it is noticeable that the majority of variation in EPV between zones is present within 30m of the opponent's try line. There is some variation 30-40m from the opponent's try line in EPV-77 (Figure 3B), which is less apparent in EPV-308 (Figure 3A) or EPV-19 (Figure 3B).

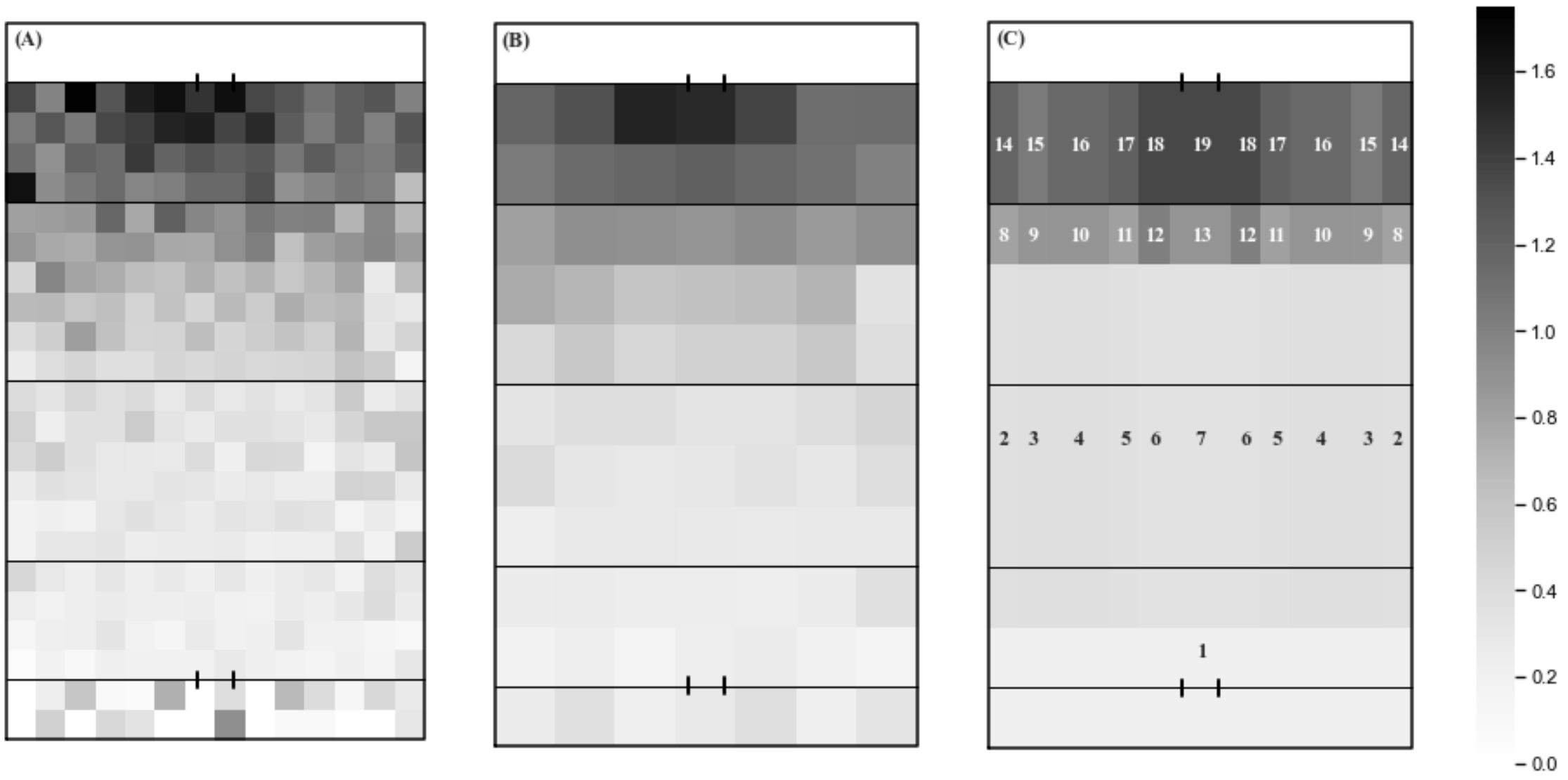

**Figure 2: EPV-308 (A), EPV-77 (B) and EPV-19 (C). Lines represent the try line, 20m line and 50m line. Within EPV-19, the numbers represent the numbers of the states as they are referenced in the text.**

*Reproducibility of attacking performances*

Figure 3 shows the reproducibility of attacking performances between fixtures for all three models. Previous attacking performances are much more reproducible in the subsequent match in the EPV-19 compared to the EPV-77 and the EPV-308. When six previous matches were considered, 95 ± 4% of subsequent matches' attacking performances were reproducible for EPV-19. Only 51 ± 12% of subsequent matches were reproducible for EPV-77; the figure was 0 ± 0% for EPV-308.

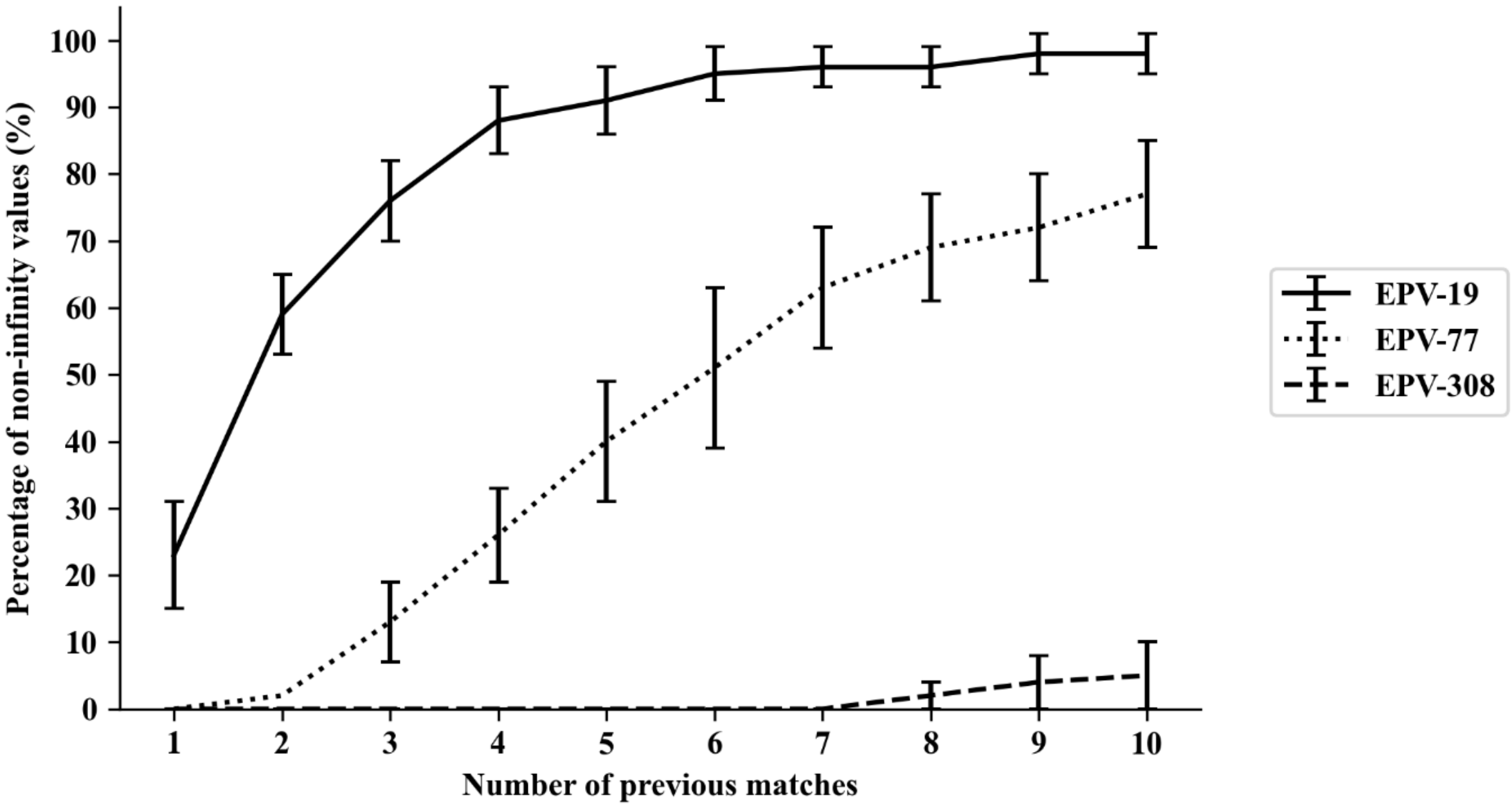

**Figure 3: Reproducibility of attacking performances between matches for EPV-308, EPV-77 and EPV-19. A higher percentage of non-infinity values indicates greater reproducibility of attacking performances. Data are mean ± SD.**

*Identifying zones important to attacking performances*

Figure 4 depicts the z-score analysis of teams' attacking performances across the 2019 Super League season. Team 4 appears to be more dependent on using wide areas (zones 3 and 4 in particular) when developing attacks 10m-70m from their own try line, whereas teams 6 and 8 appear to be much more dependent on attacking centrally (zones 5-7 for team 6, zones 5-6 for team 8) at a cost to their wide zones (zones 2-4 for team 6, zones 3-4 for team 8). Conversely, Team 9 does not appear to depend on the widest zone from 80-100m (zone 14), but their attacking is spread over the more central areas much more evenly than other teams.

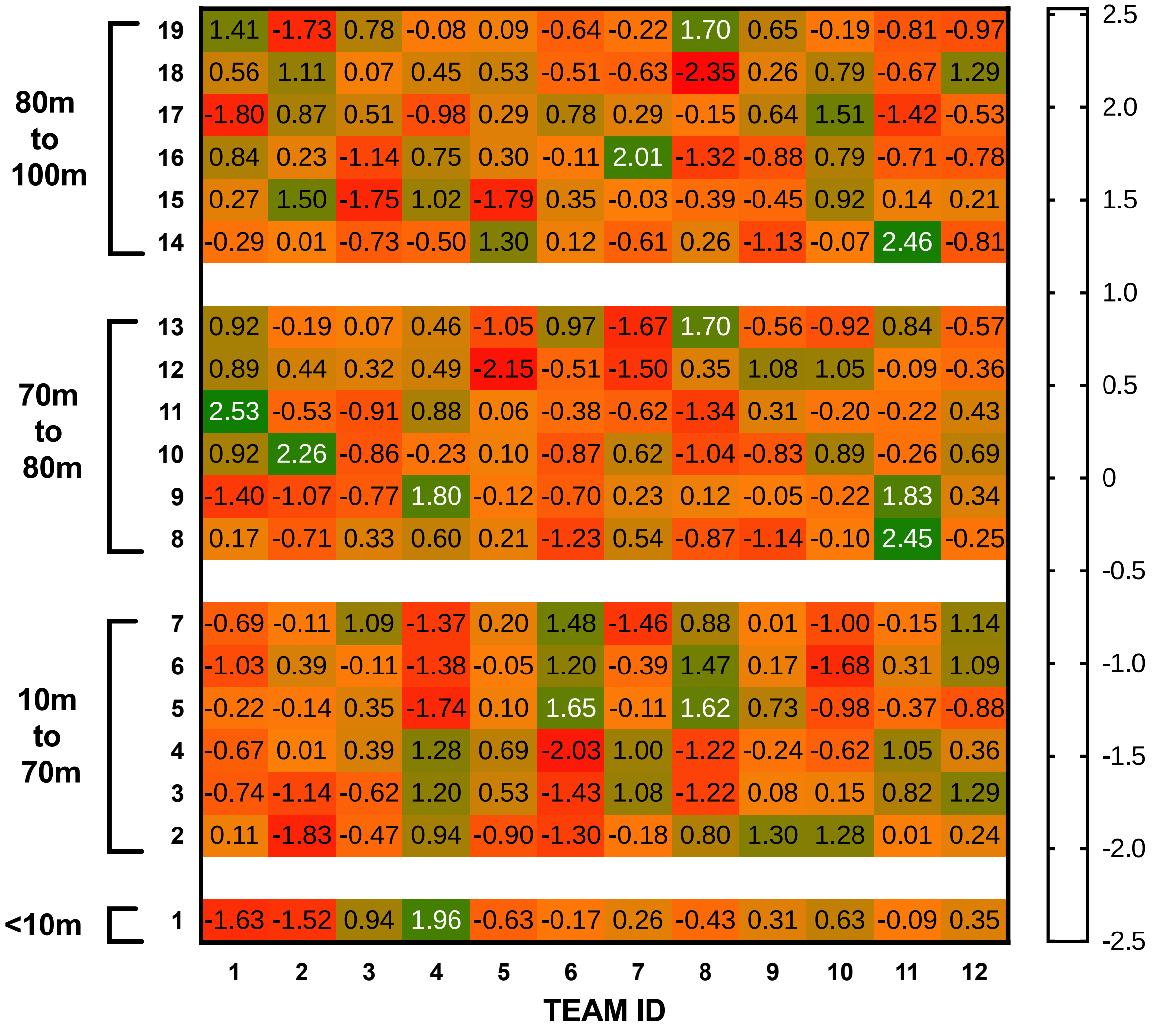

**Figure 4: Z-score of teams' attacking performances in 2019 Super League season. Numbers 1-19 reflect the zone numbers in Figure 2C. A greater value indicates a greater dependence on the zones during attacking possessions than the average team.**

## Discussion

The aims of this study were to a) generate three EPV models, two with fixed zone sizes of ~5m x 5m and ~10m x 10m (Kempton, Kennedy and Coutts, 2016), and one with aggregated zones based on the individual zones' match returns; b) identify the zone size which provided the greatest reproducibility of attacking performances between fixtures; and c) propose a framework through which an EPV model could be used in practice to evaluate a team's attacking performances. Three EPV models were produced: EPV-308, EPV-77 and EPV-19. The results show that the attacking performances of previous matches were more reproducible in subsequent matches in the EPV-19, compared to both the EPV-77 and EPV-308. Furthermore, the results showed that z-scores could be used to identify the zones a team is more or less dependent on during attacking performances.

### *Generation of EPV models*

By generating three EPV models, it is possible to compare the value that each model estimates to be generated by possessing the ball in any location on a rugby league pitch. In all three EPV models, zones were more valuable the closer they were to the opposition try line and the more central they were, which aligns with the findings of et al. (2016). Additionally, in all three models much more value is generated within 30-40m of the opposition try line, compared with >40m. This finding is similar to previous research within football, which shows that the chance of scoring is significantly reduced to below 7% when shots are taken from outside the 18 yard box (Spearman, 2018). The identification of these zones of value in all three models provides a new method through which individual possessions can be valued. Furthermore, they provide a value which tactical set plays could be measured against to establish which play may be most advantageous against a given team. However, it should be noted that within Kempton and colleagues' (2016) model, the increase in value was much more gradual from the team in possession's 20m line through to the opposition try line than in the models produced here. Whether this is due to the methodological differences (e.g. the different definition of possessions) or a difference in playing style between National Rugby League and Super League teams is unclear.

### *Reproducibility of attacking performances*

A key element of any model evaluating attacking performances is to identify how well they relate to performance in future fixtures. Despite this, few studies have attempted to evaluate this component of their models (Sarmento *et al.*, 2014). This study evaluated the reproducibility of the EPV models between fixtures via the KL Divergence. The results showed that although the EPV-308 provides significantly more variability than either the EPV-77 or EPV-19 with regards to the values of different zones, it had poor reproducibility between fixtures and therefore has limited application in practice. The EPV-19 on the other hand showed excellent reproducibility (~95% of subsequent matches) when 6 previous matches were considered. The use of six matches to analyse a team's performance provides a zone system through which the future attacking trends of an opposition rugby league team could be considered, after a relatively small number of matches.

*Identifying zones important to attacking performances*

Finally, this study outlined a framework through which the model could be used in practice to identify the zones a team was dependent on during attacking performances. Using z-score analysis to evaluate the dependence on specific zones relative to the average dependence across every team in the league provides a simple method through which important zones can be identified for individual teams. Within Figure 4, it is clear that Team 4 generates a higher proportion of its match return from different zones than Team 6 when 10m-70m from its own try line. Team 4 predominantly uses wide areas (zones 3 and 4), whereas Team 6 attacks very centrally (zones 5-7). The identification of these zones pre-match could assist teams in their tactical preparations. Furthermore, the figure shows those teams who spread their attack more evenly. For example, from 80-100m, Team 9 obtained a very small proportion of its match return widest zone (14), but they generated very similar value across the rest of the zones. It is possible that this ability to generate value across the majority of the pitch made the team difficult to defend against and could explain why they were one of the top points scorers that season.

*Limitations*

The EPV-19 provides an excellent starting point through which the tactical analysis of opposition teams can be conducted in a time efficient and easily interpretable manner in rugby league. However, it is subject to several limitations. The first of these is the use of only the start location of each play. Although it is not currently possible to provide any further information due to the limitations of the data used, it is important to note that the model could be improved if specific actions (e.g. passes, kicks or tackles) and their locations were included, alongside the locations of all players on the pitch. In soccer for example, every single action is location-coded by several providers, so a more complete model could be completed in that domain using a similar event level data only process. A second limitation of the model is that it does not analyse whether being aware of or attempting to stop an opposition team from visiting their highest valued zones during their attacking possessions is detrimental to their ability to win rugby league matches. Future studies could resolve this by building on our framework and evaluating whether there is a difference in the zones visited when a team wins or loses matches. On a related note, our study also does not attempt to directly predict future attacking trends. The authors do not consider this to be a limitation due to the variability inherent within predicting single matches, but it should be highlighted. By using the KL Divergence with the subsequent match as the true distribution, any areas with 0 values in the subsequent match are automatically believed to have 0 values in the approximating distribution. Consequently, teams could spend time preparing for the opposition to use a zone which they don't end up visiting, but they should be prepared for the vast majority of other zones that the opponent visits during attacking possessions based on the previous six matches' performances.

*Conclusion*

In conclusion, this study provides a framework through which an EPV model could be used to analyse a team's attacking performances in rugby league. The EPV-19 provides an understanding of attacking performances, which is reproducible in subsequent fixtures, when six previous matches are evaluated. Furthermore, z-score analysis comparing the proportion of match return generated by each zone relative to other teams within the league highlights the zones a team is more dependent on and therefore provides a method through which the tactical preparation of rugby league teams could be enhanced.

**Disclosure of interest**

The authors report no conflict of interest.